\documentclass[12pt]{article}
\usepackage{a4wide}

\newcommand{\be}{\begin{equation}}
\newcommand{\ee}{\end{equation}}
\newcommand{\bea}{\begin{eqnarray}}
\newcommand{\eea}{\end{eqnarray}}
\newcommand{\bref}[1]{(\ref{#1})}
\newcommand{\Lag}{\mathcal{L}}
\newcommand{\expect}[1]{\bigl<{#1}\bigr>}
\newcommand{\ms}{\ensuremath{m_\phi}}
\newcommand{\mv}{\ensuremath{m_\mathrm{X}}}
\newcommand{\mA}{\ensuremath{m_\mathrm{A}}}
\newcommand{\w}{\omega}
\newcommand{\e}{\epsilon}

\begin{document}

\title{\bf Cosmic Strings and Cooper Pairs}

\author{S. C. Davis\thanks{E-mail: S.C.Davis@swansea.ac.uk} \\ 
\em Department of Physics, University of Wales Swansea, Singleton Park, \\
\em Swansea, SA2 8PP, Wales
}

\maketitle

\begin{abstract}
It is shown that it is possible for bound fermions on a cosmic
string to form a superconducting state. Due to the attractive force
between them, particles moving in opposite directions along the string
form bound pairs. This involves a similar mechanism to superconductivity
in metals at low temperatures. The method of Gorkov is used to analyse
the system. In contrast to the situation in metals,
the unusual properties of the string fermion spectrum allow a massless
Abelian gauge field to provide the required attractive force. This
results in far stronger superconductivity than usual. A massive gauge
field can also be used, in which case the standard results apply.
\end{abstract}

\hfill SWAT/267

\section{Introduction}

It is widely believed that during its early evolution the Universe
passed through several symmetry breaking phase transitions. It is
possible that topological defects were formed at some of these
transitions. The most cosmologically significant of these are cosmic
strings~\cite{vilshel}.

As well as the strings themselves, it is also possible for particles
trapped on the strings to produce interesting cosmological effects. A
well known example of this is superconductivity~\cite{witten}. If an
electrically charged scalar field couples to the string Higgs in the
right way it will gain a non-zero expectation value in the core of the
string. Excitations of this condensate act as superconducting
currents. The currents are conserved, so the string acts as a perfect
conductor. In addition to this magnetic flux will be excluded from the
string core (the Meissner effect). This allows the string to produce
irregularities in galactic magnetic fields. One criticism of this
model is that there are no obvious candidates for the charged scalar
field.

It is also possible to use fermions as the charge carriers. If a
fermion field get its mass from the string Higgs field, then massless
states will exist in the string core~\cite{zeromodes}. As in the
bosonic case, these fermion currents are conserved, so the string is a
perfect conductor. However unlike the bosonic currents, there is no
Meissner effect, and so the string does not act as a superconducting
wire in this case (although such currents are sometimes wrongly
referred to as superconducting in the literature).

In this paper I will show that it is possible to have a truly
superconducting cosmic string with Fermi charge
carriers. This is achieved by a similar mechanism to superconductivity
in metals at low temperatures.

It is a generic feature of an interacting Fermi gas is that if there
is an {\em attractive} force between the particles, the system will have a
superconducting ground state~\cite{gorbook,fetwal}. In this state,
particles near the Fermi surface, with opposite momenta and spin will
form bound pairs, called {\em Cooper pairs}~\cite{cooper}. Formation
of these pairs lowers the energy of the system slightly, thus the
superconducting state is favoured over normal one. Bardeen, Cooper and
Schrieffer originally showed the existence of superconducting states
with a variational argument~\cite{BCS}. In this paper I will follow
the approach of Gorkov~\cite{gorkov}, in which superconductivity is
analysed using the Green's functions of the theory.

In a metal the necessary attractive force is provided by phonons. In
the cosmic string model I will consider it is provided by a massless
gauge field. If we were simply dealing with a Fermi gas, this type of
interaction would not result in a superconductivity. For example Quantum
Electrodynamics does not have a superconducting ground state. This is
due to the fact that the photon also produces a repulsive force
between like particles. However, the unusual properties of the fermion
spectrum in a cosmic string background alter this fact.

\section{Effective Two Dimensional Action}

Throughout this paper I will denote the time and position on the
cosmic string with the 2-vector $x=(t,z)$, and represent the coordinates in
the plane perpendicular to the string by $\vec x$. Similarly
$k=(k_0,k_z)$ gives the energy and momentum along the string. I shall
use the metric $g_{\mu\nu} = \mathrm{diag}(1,-1,-1,-1)$

Topologically stable cosmic strings can form at phase transitions
whose vacuum manifolds are not simply connected. The simplest example
of this is a U(1) $\rightarrow I$ symmetry breaking. If $\Phi$ is the
Higgs field responsible for the transition (with $\expect{\Phi}=v$
being the usual vacuum solution) and $X_\mu$ is the U(1) gauge field
(with coupling $g_\mathrm{X}$), then 
\be 
\expect{\Phi} = v f(r) e^{i\theta} \ , \ \  
\expect{X_\mu} = \frac{a(r)}{g_\mathrm{X}}\partial_\mu \theta 
\ee 
is a cosmic string solution. The masses of $\Phi$ and $X_\mu$ will be
denoted \ms\ and \mv. The two profile functions $f(r)$ and $a(r)$ are
zero at the string centre ($r=0$) and both tend rapidly to 1 for $r$
greater than $\ms^{-1}$ and $\mv^{-1}$ respectively.

If any fermions get their masses from $\Phi$, their spectra will have
extra states which are bound to the string. I will consider an Abelian
string model with two such Dirac fermion fields, $\Psi_\pm$, and an
additional massless gauge field $A_\mu$. At finite particle density
the Lagrangian of the system is
\bea 
&&\Lag = \sum_{s=\pm}
\bar \Psi_s \left(i\gamma^\mu D_\mu + \mu_s \gamma^0 \right)\Psi_s  
- g_+ \left[\bar \Psi_+^R \Phi \Psi_+^L 
	+ \bar \Psi_+^L \Phi^* \Psi_+^R \right] 
\nonumber \\ &&\hspace{1in}  
{}- g_-  \left[\bar \Psi_-^L \Phi \Psi_-^R 
	+ \bar \Psi_-^R \Phi^* \Psi_-^L \right]
-\frac{1}{2}\partial^\mu A^\nu \partial_\mu A_\nu +\Lag_\mathrm{string} 
\label{Lag1}
\eea
where $\Psi_s^{L,R} = \frac{1}{2}(1 \mp \gamma^5)\Psi_s$, and 
$D_\mu \Psi_s^{L,R}  = 
(\partial_\mu - ig_\mathrm{X} Q_s^{L,R} X_\mu - ig q_s A_\mu) \Psi_s^{L,R}$. 
Gauge invariance implies $Q_\pm^R - Q_\pm^L = \pm 1$. The usual masses 
of the fermions are thus $m_\pm = g_\pm v$. $\mu_\pm$ are the chemical
potentials of the two fermion species.

As well as the usual excitations of the fermion fields there are
massless states which are bound to the string~\cite{zeromodes}.  There
will also be some massive bound states. The radial dependence of the
massless states does not depend on their momentum, and so this part of
the fermion spectrum can be expressed as 
\be 
\Psi_s(x,\vec x) = \psi_s(x) u_s(r)+ (\mbox{massive states})
\label{Psi4}
\ee

If $m_\pm \ll \ms$ and $\mv \approx \ms$ then, in the chiral
representation of the Dirac matrices, the spinors $u_\pm(r)$ can be
approximated by 
\be 
u_+(r) = \left( \begin{array}{c} 1 \\ 0 \\ 0 \\ 1 \end{array}\right)  
	\sqrt{\frac{m_+}{2\pi r}} e^{-m_+ r} \ , \ \
u_-(r) = \left( \begin{array}{c} 0 \\ 1 \\ 1 \\ 0 \end{array}\right)
	\sqrt{\frac{m_-}{2\pi r}} e^{-m_- r}
\label{zmapprox}
\ee
This part of the fermion spectrum has several interesting
features. All the massless excitations of a given field have the same
spin and move in the same direction. Which direction depends on how
the field couples to $\Phi$. The spinors $u_s$ satisfy
$\gamma^0\gamma^3 u_s = s u_s$ (with $s=\pm$). The excitations of the
two fields in the above model move in opposite directions and have
opposite chirality.

We will be mainly interested in the the massless fermion states, and
so for simplicity I will ignore the massive part of the spectrum. This
will be reasonable if the particle numbers and temperature are low
enough for there to be no significant occupation of the massive
states. An effective fermion action can be obtained by integrating out
the photon field. Due to the form of the states \bref{Psi4} it is
straightforward to remove the dependence on the transverse
co-ordinates $\vec x$. The resulting effective two-dimensional fermion
action is then
\be
S_\mathrm{eff} = \int d^2x \sum_{s=\pm} 
\psi^\dagger_s \left[ i\partial_t +s i\partial_z + \mu_s\right] \psi_s
+\frac{1}{2} \sum_{s,a=\pm}\int d^2xd^2x' V^{sa}(x-x') 
\psi_s^\dagger(x) \psi_a^\dagger(x') \psi_a(x') \psi_s(x)
\label{Seff}
\ee
where the interaction between the fermions is
\be
V^{sa}(x-x') = -g^2 q_s q_a \int rdrd\theta r'dr'd\theta' \bar u_s(r)
\gamma^\mu u_s(r) \bar u_a(r') \gamma^\nu u_a(r') 
D_{\mu \nu}(x-x', \vec x-\vec x')
\label{Vdef}
\ee
$D_{\mu\nu}(x-x',\vec x-\vec x')$ is the photon propagator. When
$\Psi_\pm$ takes the form \bref{Psi4}, the $\Phi$
dependent mass terms in \bref{Lag1} cancel, so they do not feature in
\bref{Seff}. 

In a combination of momentum and position space the free-field propagator is
\be
D^{\mu\nu}_0(k,\vec x - \vec x') = g^{\mu \nu} \int \frac{k_T dk_T}{2\pi} 
\sum_l e^{il(\theta-\theta')}
J_l(k_T r) J_l(k_T r') \frac{1}{-k_0^2 + k_z^2 + k_T^2 - i\e}
\label{D0def}
\ee
The dependence on the transverse position has been expressed in polar
coordinates ($r = |\vec x|$) using a Bessel function expansion of the
plane waves.

Substituting \bref{D0def} into \bref{Vdef} gives the leading order
fermion interaction. This can be simplified using
$\gamma^0\gamma^3 u_\pm = \pm u_\pm$. Since $D^{\mu\nu}_0 \propto g^{\mu\nu}$, 
the terms in \bref{Vdef} which correspond to interactions
between particles of the same species (i.e.\ when $s=a$) cancel. Thus
$V_0^{sa}(k) = (1-\delta_{sa})V_0(k)$, with
\be
V_0(k) = -g^2 q_+ q_- 
\int \frac{1}{k_T^2 - k^2 - i\e}
\left(\int |u_+(r)|^2 J_0(k_T r) r dr\right) 
\left(\int |u_-(r')|^2 J_0(k_T r') r'dr' \right) k_T dk_T 
\label{V0def}
\ee
Since the $r$ and $r'$ integrals do not evaluate to delta functions,
transverse momentum is not conserved in this interaction. This is not
surprising since the string breaks Lorentz invariance.
If $q_+ q_- < 0$, the force between fermions of different species will
be attractive. There will be no repulsive force between like particles
since they do not interact (at least not at tree level).

An estimate of \bref{V0def} can be found using the
approximate wavefunctions \bref{zmapprox}. If $m_\pm = m$, it evaluates to 
\be
V_0(k) = U(k_z^2 -k_0^2) = -g^2 q_+ q_- \frac{2m^2}{k_z^2 - k_0^2 - 4m^2}
	\ln\left(\frac{k_z^2 -k_0^2 - i\e}{4 m^2} \right)
\label{V0}
\ee
This is the effective potential between the massless fermion states on
the string. Non-conservation of momentum changes the pole of the
propagator to a logarithmic singularity at $k_0^2=k_z^2$.
For later convenience I will define $\tilde g = g\sqrt{-q_-q_+/(2\pi)}$.

\section{Gorkov Functions and Superconductivity}
\label{sec:gorkov}

The attractive force between particles moving in opposite directions
with opposite spins in the above model suggests that the system will
have a superconducting ground state. 
We will be interested in finite temperature as well as zero
temperature effects. I will use the imaginary-time finite temperature formalism
which is easily obtained by defining $\tau = it$ and 
$k_0 = i\w_n = i(2n+1)\pi T$ (for fermions), making the replacements  
$(2\pi)^{-1} \int dk_0 \rightarrow T \sum_n$ and 
$\int dt \rightarrow \int_0^{1/T} d\tau$, and then analytically
continuing to real $\tau$ and integer $n$. 

The field equations obtained from \bref{Seff} are then
\be
\left[ -\partial_\tau +s i\partial_z + \mu_s\right] \psi_s(x) 
= -\int d^2y V(y) \psi_{-s}^\dagger(x-y) \psi_{-s}(x-y) \psi_s(x)
\label{psieq}
\ee
\be
\left[ -\partial_\tau +s i\partial_z - \mu_s\right] \psi_s^\dagger(x) 
= \int d^2y V(y)\psi_s^\dagger(x) \psi_{-s}^\dagger(x+y) \psi_{-s}(x+y) 
\label{psideq}
\ee
The usual finite temperature Green's functions are defined by
\be
G_s(x-x') = - \expect{T_\tau \psi_s(x) \psi_s^\dagger (x')}
\label{Gdef}
\ee
where $T_\tau$ indicates time ordering with respect to $\tau$ and
$\expect{\ldots}$ indicates the ensemble average. It is also necessary
to consider anomalous Green's functions~\cite{gorkov}, or Gorkov functions
\be
F_s(x-x') = \expect{T_\tau \psi_{-s}(x) \psi_s (x')}
\label{Fdef}
\ee
\be
F^\dagger_s(x-x') = -\expect{T_\tau \psi^\dagger_{-s}(x) \psi^\dagger_s (x')}
\ee
In the vacuum, or a state with a definite number of particles, $F$ will
be zero. More generally this need not be the case.

Using \bref{psieq} and \bref{psideq}, and the finite temperature
version of Wick's theorem~\cite{gorbook,fetwal}
\be
[-\partial_\tau +s i\partial_z + \mu_s] G_s (x)
+ \int d^2y V(y) \left[ G_{-s}(0^-) G_s (x) + F_s(y^+)
F_s^\dagger(x+y)\right]
= \delta(x)
\ee
\be
[-\partial_\tau -s i\partial_z - \mu_{-s}] F^\dagger_s (x)
- \int d^2y V(y) \left[G_{-s}(0^-) F^\dagger_s (x)-F^\dagger_s(y^+) G_s(x-y)
\right] = 0
\ee
here $0^\pm$ denotes the 2-vector $\lim_{\eta \to 0} (\pm \eta,0)$ with
$\eta >0$, and $y^\pm$ denotes $y + 0^\pm$. If the average number of $\psi_s$
particles per unit length is $N_s$, then 
$\expect{\psi_s^\dagger \psi_s} = G_s(0^-) =  N_s$.  

Changing to Fourier space, with
$G(k)=\int dz d\tau e^{i\w_n\tau - ik_z z}G(x)$, and similar
expressions for $F^\dagger$, and $V$,
\be
[i\w_n - \xi^s_k]G_s(k) + \Delta_s(k) F^\dagger_s(k) = 1
\label{Gkeq}
\ee
\be
[i\w_n + \xi^{-s}_{-k}]F^\dagger_s(k) + \Delta^\dagger_s(k) G_s(k) = 0
\label{Fkeq}
\ee
where $\xi^s_k = sk_z + \Sigma_s - \mu_s$. The 
self energy of the fields is $\Sigma_s = -\bar V N_{-s}$,
where \\ $\bar V = V(k=0) = \int d^2x V(x)$.

Expressing $G(0^-)$ in momentum space gives
\be
N_s = \int \frac{T dk_z}{2\pi} \sum_n e^{i\w_n\eta} G_s(k)
\label{Ndef}
\ee
which can be used to determine the chemical potentials, $\mu_s$.
The $\eta \to 0^+$ limit is implied in this and the following expressions.
As it stands, the above expression for $N$ is not
correct, since the vacuum contributions need to be subtracted off. This is
obtained by evaluating \bref{Ndef} when
$\mu$ and $T$ are zero.

The {\em gap function} $\Delta_s(k)$ is defined by
\be
\Delta_s(k) = \int \frac{T dp_z}{2\pi} \sum_m e^{i\w_m\eta} V(p-k) F_s(p)
\label{Deltadef}
\ee
It will be zero in a non-superconducting state.

For simplicity I will consider a system with equal numbers of left and
right moving particles ($N_s = N$). In
this case $\mu_s = \mu$, the Green's functions of the two particle
species are related by $G_s(k) = G(k_0, sk_z)$ and
$F_s(k) = sF(k_0,sk_z)$. Similarly $\Delta_s(k) = s\Delta(k_0,sk_z)$. The
equations \bref{Gkeq} and \bref{Fkeq} are then solved by 
\be
G(k) = -\frac{i\w_n + \xi_k}{\w_n^2 + \xi_k^2 + |\Delta(k)|^2}
\label{Gsol}
\ee
\be
F(k) = \frac{\Delta(k)}{\w_n^2 + \xi_k^2 + |\Delta(k)|^2}
\label{Fsol}
\ee
If $\Delta \neq 0$ then it can be seen from \bref{Gsol} that the there is
an energy gap in the fermion spectrum at the Fermi surface
($\xi_k=0$). This is not the same as a mass, since it is not at the
zero energy part of the spectrum. The Green's function \bref{Gsol} gives the
excitations of a state with an average of $N$ particles of each species. If 
$\Delta \neq 0$ these are not particles and holes (as in the normal ground
state), but combinations of them, called {\em quasiparticles}. 

Substituting \bref{Fsol} into \bref{Deltadef} gives an expression for
$\Delta(k)$. The only physical quantities are those involving on-shell
values of $k$, which can be found by analytic continuation to real
values of $i\w_n = k_0$. For the filled states the on-shell values of
$k_0$ are $-E_k$ with $E_k = \sqrt{\xi_k^2 + |\Delta_k|^2}$ and
$\Delta_k = \Delta(-E_k,k_z)$. The empty states have $k_0 = E'_k$
with $E_k' = \sqrt{\xi_k^2 + |\Delta'_k|^2}$ and
$\Delta'_k = \Delta(E'_k,k_z)$.

The gap equation \bref{Deltadef} for $\Delta_k$ is then
\bea
&&\Delta_k = \int \frac{d\xi_p}{2\pi} 
\frac{\Delta_p}{2E_p} n(-E_p)
U[2 E_p E_k - 2\xi_p\xi_k - |\Delta_p|^2 - |\Delta_k|^2]
\nonumber \\ && \hspace{.7in}
{}- \frac{\Delta'_p}{2E'_p} n(E'_p)
U[-2E'_p E_k - 2\xi_p\xi_k - |\Delta'_p|^2 - |\Delta_k|^2]
\label{gap1}
\eea
where $n(E) = [\exp(E/T) +1]^{-1}$. $U$ is defined by \bref{V0}. The
corresponding equation for $\Delta'_k$ can be obtained with the
replacements $\Delta_k \to \Delta'_k$ and $E_k \to -E'_k$. In the
absence of magnetic fields, we can take $\Delta_k$ to be real and
positive.  

Clearly $\Delta(k) = 0$ is a solution of \bref{gap1}. This corresponds
to the normal ground state of a Fermi system. At zero temperature this
solution has the fermion states (completely) filled, in order of
increasing energy up to the Fermi surface. It is also possible that a
$\Delta(k) \neq 0$, {\em superconducting} solution also exists. In
this case (at $T=0$) the fermion states either side of the Fermi
surface are only partially filled. Thus the Fermi surface is no longer
sharp in the superconducting state~\cite{fetwal}.

\section{The Gap Equation at Zero Temperature}

At zero temperature \bref{gap1} simplifies, and it is possible to
remove the $k$ dependence from the integral. The gap equation
can then be solved by constant $\Delta$ and $\Delta'$.
At $T=0$, $n(E) = \theta(-E)$ where $\theta$ is the step function, so
the second term of \bref{gap1} vanishes. The first can be evaluated
using the change of variables $\xi_k = \Delta \sinh 2 \kappa$ and
$\xi_p = \Delta \sinh 2\chi$. The gap equation \bref{gap1} then reduces to
\be
\Delta = \Delta \int \frac{d\chi}{2\pi} U[4 \Delta^2 \sinh^2 (\chi-\kappa)]
\label{D-eq1}
\ee
Similarly, using $\xi_k = -|\Delta'| \sinh 2 \kappa$, the
corresponding equation for $\Delta'$ becomes
\be
\Delta' = 
\Delta\int \frac{d\chi}{2\pi} U[-(|\Delta'| -\Delta)^2 -4|\Delta'|\Delta
\cosh^2(\chi-\kappa)]
\label{D+eq1}
\ee
The dependence on $\kappa$ (and hence $k$) is then simple to
remove. Unfortunately analytic evaluation of the resulting integrals does not
seem possible. However we expect $\Delta$ to be very small. This fact
can be used to approximate the integrands of
(\ref{D-eq1},\ref{D+eq1}). For simplicity I will take $m_+ = m_- =m$,
in which case $U[-(p-k)^2]$ is given by \bref{V0}. The main
contribution to the integrals occurs for moderate values of $\chi$, when
$e^{-|\chi|} \ll e^{|\chi|} \ll 2 m/\Delta$. In this region the denominator of
\bref{V0} is approximately constant and $2\sinh\chi \approx
e^{|\chi|}$. For $|\chi| > \ln(2m/\Delta)$ the integrand of
\bref{D-eq1} tends to zero rapidly. This suggests the approximation
\be
\frac{1}{2\pi}U[(2\Delta\sinh\chi)^2] \approx
\tilde g^2 \left[\ln \frac{2m}{\Delta} - |\chi|\right] 
\theta\left(\ln \frac{2m}{\Delta} - |\chi|\right)
\label{app1}
\ee
Using this approximation \bref{D-eq1} becomes
\be
\Delta \approx \Delta \tilde g^2\int_{-\ln(2m/\Delta)}^{\ln(2m/\Delta)}
\left(\ln \frac{2m}{\Delta} - |\chi| \right) d\chi
= \Delta \tilde g^2\left(\ln \frac{2m}{\Delta} \right)^2
\ee
Thus either $\Delta = 0$ (normal solution) or 
\be
\Delta \approx 2m \exp\left(-\frac{1}{\tilde g}\right) 
= 2m \exp\left(-\sqrt{\frac{2\pi}{-q_+q_-}} \frac{1}{g}\right) 
\label{Del1}
\ee
which is the superconducting solution. This value of $\Delta$
contrasts with the more usual result of $\Delta \sim \exp(-1/g^2)$. It
can be seen that \bref{Del1} has an essential singularity at
$g=0$. This is a generic feature of superconductors, and explains why
they cannot be analysed with standard perturbation theory.

To leading order the approximation \bref{app1} also works for
\bref{D+eq1}, and so $\Delta' \approx \Delta$. There will be a
small phase difference between the two constants because the right-hand
side of \bref{D+eq1} includes a small imaginary part.

The total energy of the state is given by
\be
E = \lim_{x \to 0^-} \sum_{s=\pm}[si\partial_z + \Sigma_s]G_s(x)
\label{Edef1}
\ee
When there are equal numbers of $\psi_+$ and $\psi_-$ particles, the
momentum space version of \bref{Edef1} reduces to
\be
E = 2\int \frac{T dk_z}{2\pi} \sum_n e^{i\w_n\eta} (\xi_k + \mu)G(k)
\label{Edef}
\ee
where $G(k)$ is given by \bref{Gsol}. Using \bref{Ndef}, the chemical
potential is found to be  $\mu = 2\pi N - \bar V N$. As it
stands, for $\Delta \neq 0$, \bref{Edef} is divergent. If we introduce
a cutoff at $|\xi_k| = \Lambda$ in the divergent part of the integral,
it evaluates to
\be
E = 2 \pi N^2 - 2\bar V N - \frac{1}{4\pi}\Delta^2
\left[1 + \ln\frac{4\Lambda^2}{\Delta^2}\right]
\ee

In the calculations of section~\ref{sec:gorkov} all energies and
momenta were assumed to contribute to the superconducting
condensate. In practice contributions from the massive part of the
fermion spectrum (which were ignored for simplicity) will alter the
effective potential for $|\xi_k| > m$, producing some sort of cutoff,
probably around the off-string fermion mass.  This suggests that $m$
is a reasonable value for $\Lambda$. In this case
\be
E = 2 \pi N^2 - 2\bar V N 
- \frac{1}{4\pi} \Delta^2\left[1+\frac{2}{\tilde g}\right]
\label{energy}
\ee
Clearly (whatever the value of $\Lambda$), \bref{energy} is smaller
when $\Delta \neq 0$, thus the superconducting ground state is
favoured over the normal $\Delta=0$ one.

Superconducting cosmic string models have been studied previously
using a fundamental bosonic scalar field as a superconducting
condensate~\cite{witten}. A connection with this work can be made by
considering an effective bosonic field $\psi_+ \psi_-$ (not to be
confused with $\psi \psi^\dagger$). This will have a non-zero
expectation value in the superconducting state. This is equal to
$F(x \to 0)$. Evaluating $\expect{\psi_- \psi_+}$ with
\bref{Fsol} gives a divergent integral.
As with \bref{energy}, it is reasonable to impose a cutoff. In this case 
$\expect{\psi_- \psi_+} \sim \Delta / \tilde g$, which is very small.

\section{Critical Temperature}

At high temperatures, thermal excitations prevent the fermions forming
the bound pairs required for the superconducting state. As the
temperature is increased from zero, a progressively smaller fraction
of the fermions will be paired up. At some critical temperature $T_c$ there
will be no Cooper pairs left and the system will just be a normal
(i.e\ non-superconducting) gas at finite temperature.

Solving \bref{gap1} when $T \neq 0$ is much harder than the zero
temperature case. It is no longer possible to remove the $k$
dependence from the integral, and $\Delta$ is no longer a
constant. Despite this it is still possible to estimate the critical
temperature, $T_c$, above which the ground state of the system is no
longer superconducting. This can be found by finding the value of $T$
which satisfies \bref{gap1} when $\Delta \to 0$.

While taking $\Delta$ small simplifies \bref{gap1}, the $k$ dependence
of the integral still creates problems. To estimate $T_c$ I will take
$\Delta_k$ and $\Delta'_k$ to be equal and constant. Using the
approximations in the previous section, the gap equation
when $\xi_k = 0$ reduces to
\be
1 \approx \tilde g^2 \int_{-\ln(2m/\Delta)}^{\ln(2m/\Delta)} d\chi
\left(\ln \frac{2m}{\Delta} - |\chi| \right) 
\tanh \left(\frac{\Delta}{2T_c} e^{|\chi|}\right)
\label{Tc1}
\ee
where $\xi_p = \Delta \sinh 2\chi$. Changing variables to 
$w= \Delta/(2T_c) e^\chi$, \bref{Tc1} becomes
\be
\frac{1}{\tilde g^2} \approx \int_0^{m/T_c} \ln \left(\frac{T_c}{m} w\right)
\frac{\tanh w}{w} dw
\ee
The leading order (in $T_c/m$) contribution of the integral can be found
by integration by parts
\be
\frac{1}{\tilde g^2}  \sim - \left(\ln \frac{m}{T_c}\right)^2
\label{Tc2}
\ee
 Solving this gives 
\be
T_c \sim m \exp\left(-\frac{1}{\tilde g}\right) 
\sim \Delta|_{T=0}
\ee
For $\xi_k \neq 0$ there will be additional terms of order
$\ln m/T_c$ on the right hand side of \bref{Tc2}. Since $m \gg T_c$,
these can be neglected to first order and so ignoring the $k$
dependence of $\Delta$ is a reasonable approximation.
 
The corresponding result for metals is also $T_c \sim \Delta$,
although in this case $\Delta$ is far smaller. Since $T_c \ll m$, the
cosmic string will not become superconducting until the universe has
cooled considerably. The temperature at which this occurs is extremely
sensitive to $g$. For example, suppose that $\Psi_\pm$ are GUT scale
fermions with $m \sim 10^{16}$GeV. If $\tilde g = 1/10$ then 
$T_c \sim 10^{11}$GeV, while if $\tilde g = 1/100$, $T_c \sim 10^{-28}$GeV. 
It is therefore possible to construct a model which
becomes superconducting at a temperature far below the string
formation scale, without having to introduce unnaturally small
fundamental constants. Note that it is still possible to get the
standard non-superconducting fermion currents at any temperature. All
that is required is different numbers of left and right moving
particles on the string.

\section{Higher Order Effects}

So far I have ignored higher order corrections to the
interaction. These will be particularly significant for calculating
the self energy of the fermions. To leading order this is proportional
to $\bar V = V^{+-}(k=0)$, which according to \bref{V0} is divergent.

Taking into account loop corrections from the fermion states on the
string the potential $V$ is given by
\be
V^{sc} = V^{sc}_0 + \sum_{a,b=\pm} V^{sa}_0 \Pi_{ab}V^{bc} 
\label{Vcor}
\ee
where $V^{sc}_0$ is the free field potential \bref{V0def} and $\Pi_{sr}$ is the
polarisation. To leading order in $g$
\be
\Pi_{ab}(k) = -\delta_{ab} \int \frac{d^2p}{(2\pi)^2} G_a(k+p) G_a(p) 
\label{Pidef}
\ee
At zero temperature the contribution from the massive fermion states
will be the same as in the vacuum. Since (after renormalisation) these
do not contribute to $V$ when $k=0$ they can be ignored.

To leading order in $g$, we can take $\Delta=0$ in \bref{Pidef}. In
this case the Green's functions reduce to $G(k) = (i\w_n - \xi_k)^{-1}$. 
There are also contributions to \bref{Pidef} involving
$F$ and $F^\dagger$, but since they are $O(\Delta^2)$ they can also be
dropped. With these approximations \bref{Pidef} evaluates to
\be
\Pi^{ab}(k) = \frac{1}{2\pi}\delta_{ab}
\ee
at any temperature. Substituting this into \bref{Vcor} and solving gives
\be
V^{+-} = \frac{(2\pi)^2 V_0}{(2\pi)^2 -V_0^2}
\ , \ \ \ \ 
V^{ss} = \frac{2\pi V^2_0}{(2\pi)^2 -V_0^2}
\ee
Thus $V^{+-} \to 0$ as $k \to 0$, so the terms considered in
section~\ref{sec:gorkov} do not contribute to the self energy of the fermion
fields at all.

The corrections to $V$ also create a pole at 
$k_z^2-k_0^2 = 4m^2 e^{-1/\tilde g^2}$. The residue is of order
$\tilde g^{-2} e^{-1/\tilde g^2} \ll \Delta$, and so the pole does not
significantly alter the results of the previous sections. 

In addition it can be seen that the interaction between fermions of
the same species is no longer zero. This will lead to further
corrections to the self energy of the fermions. To fully analyse this,
Green's functions of the form 
\be
G_{s,-s}(x-x') = -\expect{T_\tau \psi_s(x) \psi_{-s}^\dagger(x')}
\label{altGdef}
\ee
need to be considered in addition to the usual Green functions
\bref{Gdef}. The main contribution to $G_{s,-s}$ will come from
particle-antiparticle interactions with energies near to zero, which
may give the Green's functions an effective mass term. The
superconducting effects result from quasiparticle interactions with
energies near the Fermi surface i.e.\ in a different part of the
spectrum. The corrections arising from \bref{altGdef} are therefore
unlikely to affect the superconductivity of the system and I will
not consider them further in this paper.

It is also possible that the photon could gain a mass, perhaps from
some other fields not bound to the string, or from the massive $\Psi$
states when $T \neq 0$. In this case a mass term will appear in
\bref{D0def}. Alternatively the the force could be provided by a
massive boson, possibly the string gauge field $X$. If $X$ were used
there would be contributions to the potential from string bound states
as well as the continuum states. For a constant mass, the
potential will resemble \bref{V0def} or \bref{V0}, with $k_z^2$
replaced by $k_z^2 + \mA^2$. In this case the self energy of the particles is 
\be
\Sigma = - 2\pi N \tilde g^2 \left(\frac{\mA^2}{4m^2} - 1\right)^{-1}
	\ln\frac{\mA}{2m}
\ee

If $\mA \gg \Delta$, \bref{app1} is not a good approximation of $V$
and
\be
\frac{1}{2\pi}U[\mA^2 + (2\Delta\sinh\chi)^2] \approx
\frac{\tilde g^2}{2} 
\frac{\ln \left(\frac{\mA}{2m}\right)^2}{\left(\frac{\mA}{2m}\right)^2 - 1}
\theta\left(\ln \frac{\mA}{\Delta} - |\chi|\right)
\ee
is more appropriate. In this case the non-zero solution of the gap equation is
\be
\Delta \approx \mA \exp\left(- \frac{\left(\frac{\mA}{2m}\right)^2-1}
{\ln \left(\frac{\mA}{2m}\right)^2}\frac{1}{\tilde g^2}\right)
\label{Del2}
\ee
which resembles the more usual result for metals.
Strictly speaking, if $A$ is massive the cosmic string is no longer
superconducting, since the fermions no longer carry (unbroken)
charge. However $\Psi_\pm$ could carry other charges in which case the
system will superconduct. Of course this assumes that interaction with
the other gauge bosons is not strong enough to cancel the attractive
force arising from the field $A$.

\section{Conclusions}

It has been shown that it is possible for a cosmic string with
massless fermion bound states to have a superconducting ground
state. In order for this to happen there must be fermions moving in
both directions along the string, with an attractive force between
them.

If the oppositely moving fermions have opposite charges, the
attractive force can be provided by a gauge boson with zero
mass. This is in contrast to an interacting Fermi gas in a spatially
homogeneous (i.e.\ no cosmic string) background. Metals at low temperature
fall into this category. In this case a massless intermediate boson
will not give a superconducting ground state.
The unusual features of the string fermion spectrum remove the
repulsive part of the interaction, which would normally
prevent superconductivity.

One characteristic of a superconducting state is an energy gap
$\Delta$ in the excitations of the state. The occurs at the Fermi
surface, and is usually of order $\exp(-1/g^2)$, where $g$ is coupling
strength of the gauge boson providing the force between the
fermions. If the gauge boson is massless and the fermions are bound to
a cosmic string, $\Delta$ is far higher, of order $\exp(-1/g)$. This is
lower than might be expected, since non-conservation of momentum by the
string reduces the force.

The quantity $\Delta$ also determines the critical temperature of the
system, above which it ceases to be superconducting. Since $\Delta$ is
very small, this will be far lower than the fundamental energy scale
of the model. 

The electromagnetic effects of the superconductor are proportional 
to the number of superconducting fermions in the system. At zero
temperature this is all of them, so the electromagnetic effects need
not be small even though $\Delta$ is. These can be analysed by, for
example, perturbing the Green's functions~\cite{gorbook} (if the field
is weak).

Some possible candidates for the fermions to make up the condensate
are superheavy GUT fields, such as those occurring in $E_6$ or
SO(10).

$E_6$ has superheavy versions of electrons, neutrinos and down
quarks. An attractive force could be provided by GUT
interactions. Since these would involve massive bosons, $\Delta$ would
be given by \bref{Del2} rather than \bref{Del1}. Alternatively if the
force were provided by $W$-bosons, the massless boson result
\bref{Del1} would apply, at least until the electroweak phase
transition.

The only fermions coupling to the string Higgs field in an SO(10) GUT are
right-handed neutrinos. Since at least two species are required to
form Cooper pairs, this type of GUT string does not have
superconducting states.  

Interestingly, electroweak strings have $u$, $d$ and $e$ massless
string bound states. Conveniently the up quarks (positively charged)
move in the opposite direction to the electrons and down quarks
(negatively charged), so these strings have all the required features
of the model discussed in this paper. Unfortunately electroweak
strings do not appear to be stable. However the effects of the
fermion fields have not been fully investigated in these models, so
they could be stable under the right conditions.

\section*{Acknowledgements}

I wish to thank Warren Perkins and Simon Hands for useful discussions,
and PPARC for financial support.

\providecommand{\pr}[3]{{Phys.\ Rev. \ }{\bf #1}, #3, (#2)}
\providecommand{\npb}[3]{{Nucl.\ Phys.\ }{\bf B #1}, #3, (#2)}
\providecommand{\jetp}[3]{{Sov.\ Phys.\ JETP }{\bf #1}, #3 (#2)}
\providecommand{\zetf}[3]{{J.\ Exptl.\ Theoret.\ Phys.\ }{\bf #1}, #3 (#2)}

\end{document}